\newcommand{\jhep}{0}

\ifnum \jhep=1
\documentclass[11pt,a4paper]{article}
\else
\documentclass[10pt,english,notitlepage,10pt]{revtex4-1}
\fi

\ifnum \jhep=1
\usepackage{jheppub}
\fi

\usepackage[T1]{fontenc}
\usepackage[latin9]{inputenc}
\usepackage{amsmath}
\usepackage{amssymb}
\usepackage{esint}
\usepackage{graphicx}
\usepackage{caption}
\usepackage{subcaption}
\usepackage{bm}
\usepackage{color}

\makeatletter

\@ifundefined{textcolor}{}
{%
    \definecolor{BLACK}{gray}{0}
    \definecolor{WHITE}{gray}{1}
    \definecolor{RED}{rgb}{1,0,0}
    \definecolor{GREEN}{rgb}{0,1,0}
    \definecolor{BLUE}{rgb}{0,0,1}
    \definecolor{CYAN}{cmyk}{1,0,0,0}
    \definecolor{MAGENTA}{cmyk}{0,1,0,0}
    \definecolor{YELLOW}{cmyk}{0,0,1,0}
}

\makeatother

\definecolor{fix}{rgb}{0,0,0}

\ifnum \jhep=1
\title{Extended DBI massive gravity with generalized fiducial metric}

\author[a,b]{Tossaporn Chullaphan,}
\author[c]{Lunchakorn Tannukij}
\author[a,d]{and Pitayuth Wongjun}

\affiliation[a]{The Institute for Fundamental Study, Naresuan University, Phitsanulok 65000, Thailand}
\affiliation[b]{Department of Physics, Faculty of Science, Udon Thani Rajabhat University, Udon Thani 41000, Thailand}
\affiliation[c]{Department of Physics, Faculty of Science,  Mahidol University, Bangkok 10400, Thailand}
\affiliation[d]{Thailand Center of Excellence in Physics, Ministry of Education,
Bangkok 10400, Thailand}

\emailAdd{chullaphan.t@gmail.com}
\emailAdd{l\_tannukij@hotmail.com}
\emailAdd{pitayuthw@nu.ac.th}

\abstract{
We consider an extended model of DBI massive gravity by generalizing the fiducial metric to be an induced metric on the brane corresponding to a domain wall moving in five-dimensional Schwarzschild-Anti-de Sitter spacetime. The model admits all solutions of FLRW metric including flat, closed and open geometries while the original one does not. The background solutions can be divided into two branches namely self-accelerating branch and normal branch. For the self-accelerating branch, the graviton mass plays the role of cosmological constant to drive the late-time acceleration of the universe. It is found that the number degrees of freedom of gravitational sector is not correct similar to the original DBI massive gravity. There are only two propagating degrees of freedom from tensor modes. For normal branch, we restrict our attention to a particular class of the solutions which provides an accelerated expansion of the universe. It is found that the number of degrees of freedom in the model is correct. However, at least one of them is ghost degree of freedom which always present at small scale implying that the theory is not stable.}

\arxivnumber{1502.08018}

\fi

\begin{document}

\ifnum \jhep=0
\title{Extended DBI massive gravity with generalized fiducial metric}

\author{Tossaporn Chullaphan\footnote{Email: chullaphan.t@gmail.com}}

\affiliation{The Institute for Fundamental Study, Naresuan University,
Phitsanulok 65000, Thailand}

\affiliation{Department of Physics, Faculty of Science, Udon Thani Rajabhat University, Udon Thani 41000, Thailand}

\author{Lunchakorn Tannukij \footnote{Email: l\_tannukij@hotmail.com}}

\affiliation{ Department of Physics, Faculty of Science,  Mahidol University, Bangkok 10400, Thailand}

\author{Pitayuth Wongjun \footnote{Email: pitayuthw@nu.ac.th}}

\affiliation{The Institute for Fundamental Study, Naresuan University,
Phitsanulok 65000, Thailand}

\affiliation{Thailand Center of Excellence in Physics, Ministry of Education,
Bangkok 10400, Thailand}


\begin{abstract}
We consider an extended model of DBI massive gravity by generalizing the fiducial metric to be an induced metric on the brane corresponding to a domain wall moving in five-dimensional Schwarzschild-Anti-de Sitter spacetime. The model admits all solutions of FLRW metric including flat, closed and open geometries while the original one does not. The background solutions can be divided into two branches namely self-accelerating branch and normal branch. For the self-accelerating branch, the graviton mass plays the role of cosmological constant to drive the late-time acceleration of the universe. It is found that the number degrees of freedom of gravitational sector is not correct similar to the original DBI massive gravity. There are only two propagating degrees of freedom from tensor modes. For normal branch, we restrict our attention to a particular class of the solutions which provides an accelerated expansion of the universe. It is found that the number of degrees of freedom in the model is correct. However, at least one of them is ghost degree of freedom which always present at small scale implying that the theory is not stable.
\end{abstract}

\fi

\maketitle 
\flushbottom

\section{Introduction}

Massive gravity is a theory of a massive spin-2 graviton, a generalization from the Einstein gravity which corresponds to the massless graviton. The linearized massive gravity was constructed by Fierz and Pauli \cite{Fierz:1939ix} by adding an interaction term into the linearlized Einstein-Hilbert action.
However, the Fierz-Pauli theory suffers from the disagreement between the predictions made from the massless limit of the theory and those made from the general relativity, which was proposed by van Dam, Veltman, and Zakharov (known as vDVZ discontinuity) \cite{vanDam:1970ab,Zakharov:1970cd}. The solution to the discontinuity was clarified by Vainshtein \cite{Vainshtein:1972sx}, that the linear approximation breaks down for the massless limit of the theory and then such discontinuity can be lifted in non-linear theory of massive gravity. The cost of the non-linear generalization is the existence of the ghostly sixth degree of freedom, the Boulware-Deser (BD) ghost \cite{Boulware:1973my}. The ghost free generalization of massive gravity, the so-called dRGT massive gravity, was successfully constructed by de Rham, Gabadadze, and Tolley \cite{deRham:2010ik,deRham:2010kj} which the suitable interaction terms are chosen such that there exists no BD ghost in the theory.

In order to investigate the cosmological implication of the dRGT massive gravity, it is convenient to use the Friedman-Lema\^itre-Robertson-Walker (FLRW) metric as the physical metric. It was found that the dRGT massive gravity does not admit flat and closed FLRW solution \cite{D'Amico:2011jj,Gumrukcuoglu:2011ew} while the self-accelerating open FLRW solutions were found in \cite{Gumrukcuoglu:2011ew}. A more general issue on the model was investigated by replacing the Minkowski fiducial metric with a de Sitter or FLRW fiducial metric \cite{Fasiello:2012rw,Langlois:2012hk,Gumrukcuoglu:2011zh,Langlois:2013cya}. It was found that the model admits not only open but also flat and closed FLRW solutions and one of the solutions provides the self-accelerating expansion of the late-time universe.

Although the dRGT theory succeeds in providing the self-accelerating solution for the universe, the number of propagating degrees of freedom in the theory is not correct for the cosmological solution. Basically in four-dimensional spacetime, there are 5 degrees of freedom for the massive gravity while there are only 2 propagating degrees of freedom if the cosmological Friedman-Lemaitre-Robertson-Walker (FLRW) solution is assumed \cite{Gumrukcuoglu:2011zh}. Moreover, De Felice, Gumrukcuoglu and Mukohyama \cite{DeFelice:2012ac} found that in the FLRW limit of the anisotropic solutions of the dRGT massive gravity, there is a non-BD ghost instability. This suggests the motivation for modifications on the dRGT massive gravity in order to cover the viable cosmological solutions.

One of possibilities to obtain a viable cosmological model of massive gravity is that, one may generalize the dRGT theory by breaking the isotropy and study the anisotropic solution \cite{Mukohyama:2012op, Antonio:2013qr}. An alternative possibility can be obtained by introducing new degrees of freedom along with its coupling to the massive graviton. For example, one can promote the graviton mass to be a function of an extra scalar field, the so-called mass-varying massive gravity \cite{Huang:2012pe}. One can also multiply a fiducial metric with the conformal factor depending on an extra scalar field \cite{D'Amico:2012zv}. The action of this extended dRGT massive gravity is invariant under quasidilaton global symmetry and then known as the quasidilaton dRGT massive gravity. Unfortunately, It was found that  the self-accelerating solutions of the model are always plagued by ghost instability \cite{Gumrukcuoglu:2013nza,D'Amico:2013kya}. Thus, the quasidilaton massive gravity is further extended in order to avoid the ghost instability by introducing a coupling between the quasidilaton scalar field and St$\ddot{\text{u}}$ckelberg fields while the quasidilaton symmetry is still preserved \cite{DeFelice:2013za,DeFelice:2013gm,Heisenberg:2015voa}. It was also found that this extension can provide the correct number {\color{fix}of} degree{\color{fix}s} of freedom in the theory. It also was found that the effect of gravitational wave in quasidilaton massive gravity can be made larger {\color{fix}than} the Hubble parameter leading to an explanation of the suppression of the power spectrum in cosmic microwave background \cite{Kahniashvili:2014wua}. Recently, another possible way to obtain viable model of massive gravity is that the extension of dRGT massive gravity by introducing nontrivial coupling to the matter field \cite{Gumrukcuoglu:2014xba,Solomon:2014iwa}.

Another interesting extended massive gravity theory is Dirac-Born-Infeld (DBI) massive gravity \cite{Hinterbichler:2013dv}. This extension is obtained by introducing DBI scalar field into the theory through the coupling term in such a way that the scalar possess{\color{fix}es} generalized Galileon shift symmetries \cite{Gabadadze:2012tr}. Like original dRGT massive gravity, it was shown that the model does not admit flat and closed FLRW solutions as well as the number of degree{\color{fix}s} of freedom is still not correct \cite{Andrews:2013uca}. This may occur from using the Minkowski fiducial metric. Similarly to the original dRGT massive gravity, one may consider the generalized version of the fiducial metric in order to obtain the solutions which admit the flat and closed FLRW universe. Since the flat Minkowski fiducial metric in DBI massive gravity is motivated from the induced metric from brane world scenario, it is worthwhile {\color{fix}to put a more general induced metric in consideration.} In this work, we consider the extended dRGT massive gravity in which the fiducial metric is obtained from the induced metric of the brane moving in five-dimensional Schwarzschild-Anti-de-Sitter (Schwarzschild-AdS) spacetime \cite{Mukohyama:1999wi}. It is found that the model not only admits the open but also flat and closed FLRW universe. The background solutions can be divided into two branches; a self-accelerating branch and a normal branch. For the self-accelerating branch, the graviton mass will play the role of the cosmological constant to drive the late-time expansion of the universe. For the normal branch, the general solutions are complicated and it is not easy to obtain the analytical solutions. {\color{fix}To extract some information of this branch, we} consider a {\color{fix}specific} class of the solutions and found that they can also provide the late-time expansion of the universe. The cosmological perturbations of the model are expanded around these background solutions. The results of cosmological perturbations in flat FLRW universe is that there are no ghost in the linear level while the number degree of freedom is still not correct in the self-accelerating branch. For the normal branch, there has the correct number {\color{fix}of} degree{\color{fix}s} of freedom but the ghost instability always presents in the theory, at least at small scale. This problem may be alleviated by introducing the coupling term similar to one in the generalized quasidilaton massive gravity.

The paper is organized as follows: In Section \ref{sec:eom}, we present the setup of the model and find the background solutions. In Section \ref{sec:pert} the cosmological perturbations are analyzed in order to find whether the number of the degrees of freedom  in the theory is correct and whether there is a ghost among them. Finally, we summarize the results in Section \ref{sec:con}.

\section{The model and the background equations\label{sec:eom}}
In this section, we will consider dRGT massive gravity including DBI Galileon term as follows
\begin{eqnarray}
S &=& S_{dRGT} + S_{\text{DBI}},\nonumber
\\
&=& \frac{M_{Pl}^2}{2} \int d^4 x \sqrt{-g} \left[ R[g] + 2 m^2_g (\mathcal{L}_2[g,\bar{g}] +\alpha_3\mathcal{L}_3[g,\bar{g}] +\alpha_4\mathcal{L}_4[g,\bar{g}] ) \right] \nonumber
\\
&&\qquad- \Lambda^4 \int d^4 x \sqrt{-\bar{g}}, \label{action1}
\end{eqnarray}
where $g_{\mu\nu}$ is the physical metric which defines the measurement on the spacetime and $\bar{g}_{\mu\nu}$ is the fiducial metric introduced to construct nontrivial nonlinear interaction terms in massive gravity. The first term in the action, the Einstein-Hilbert term corresponds to the Einstein's general relativity. The second term is the interaction terms or mass terms, involving both the physical and the fiducial metric,
characterized by 3 parameters; $m_g$, $\alpha_3$, $\alpha_4$, where $m_g$ can be interpreted as a mass of graviton.
The interaction terms are constructed, to avoid BD ghost, as follows \cite{deRham:2010kj},
\begin{eqnarray}
\mathcal{L}_2[g,\bar{g}]&=&\frac{1}{2}\left([\mathcal{K}]^2-[\mathcal{K}^2]\right),\\
\mathcal{L}_3[g,\bar{g}]&=&\frac{1}{3!}\left([\mathcal{K}]^3-3[\mathcal{K}][\mathcal{K}^2]+2[\mathcal{K}^3]\right),\\
\mathcal{L}_4[g,\bar{g}]&=&\frac{1}{4!}\left([\mathcal{K}]^4-6[\mathcal{K}]^2[\mathcal{K}^2]+3[\mathcal{K}^2]^2+8[\mathcal{K}][\mathcal{K}^3]-6[\mathcal{K}^4]\right),
\end{eqnarray}
where the square bracket denotes the trace operation with respect to the physical metric and the building-block tensor is defined as
\begin{eqnarray}
{\mathcal{K}^\mu}_\nu = \delta^\mu_\nu-{\left(\sqrt{g^{-1}\bar{g}}\right)^\mu}_\nu.
\end{eqnarray}
The square root denotes the tensor which upon being squared equals the ${(g^{-1}\bar{g})^\mu}_\nu$. Moreover, the fiducial metric $\bar{g}_{\mu\nu}$ is an induced metric from the five-dimensional fiducial metric $\tilde{g}_{AB}$,
\begin{eqnarray}
\bar{g}_{\mu\nu} = \partial_\mu X^A \partial_\nu X^B \tilde{g}_{AB},
\end{eqnarray}
where {\color{fix}all of $X^A=(\varphi^0,\varphi^1,\varphi^2,\varphi^3,X^5)$}, {\color{fix} the so-called St\"{u}ckelberg fields}, transform as scalars introduced to restore general covariance of the theory. Lastly, the last and extension term is a leading order term of the action which is invariant under the Galileon shift transformation, also known as DBI action, where $\Lambda$ is interpreted as a tension on the brane.

For the physical metric, we will consider the Friedmann-Lema\^itre-Robertson-Walker (FLRW) metric,
\begin{eqnarray}
ds^2 = -N^2(t) dt^2 + a^2(t) \Omega_{ij}(x^k) dx^i dx^j.
\end{eqnarray}
Here, the {\color{fix}l}atin indices run over all the spatial indices; 1, 2, and 3. The 3-dimensional tensor $\Omega_{ij}$ is a 3-space metric defined by
\begin{eqnarray}
\Omega_{ij}(\varphi^k)=\delta_{ij}+\frac{\kappa\delta_{il}\delta_{jm}\varphi^l\varphi^m}{1-\kappa\delta_{lm}\varphi^l\varphi^m}.
\end{eqnarray}
Here, the curvature of the 3-space is defined by the value of $\kappa$ where the closed, flat, and open geometry correspond to $\kappa$ being positive, zero, and negative respectively.

The dRGT massive gravity with flat Minkowski fiducial metric including DBI Galileon term have been investigated in \cite{Hinterbichler:2013dv,Gabadadze:2012tr,Andrews:2013uca}. It was shown that the model does not admit flat and closed FLRW solutions as well as the number of degree{\color{fix}s} of freedom is still not correct. These behaviors are similar to the dRGT massive gravity. In order to obtain the flat and closed FLRW solutions, one may generalize the fiducial metric from the Minkowski one. The generalizations in order to obtain the flat and closed FLRW solutions in dRGT massive gravity were investigated by using both FLRW and de Sitter fiducial metric \cite{Fasiello:2012rw,Langlois:2012hk, Gumrukcuoglu:2011zh, Langlois:2013cya}. For the DBI massive gravity model, the fiducial metric can be interpreted as {\color{fix}an induced} metric {\color{fix}in the} brane world scenario. In order to generalize the fiducial metric to obtain the closed and flat solutions for the DBI massive gravity, we consider the form of the fiducial metric, $\tilde{g}_{AB}$, which corresponds to a domain wall moving in five-dimensional Schwarzshild-AdS spacetime \cite{Mukohyama:1999wi},
\begin{eqnarray}
ds^2 = -f(X^5) d T^2 + (X^5)^2 \Omega_{ij}(\varphi^k) d\varphi^i d\varphi^j + \frac{1}{f(X^5)}(dX^5)^2,
\end{eqnarray}
where
\begin{eqnarray}
f(X^5) = \kappa - \frac{\mu}{(X^5)^2} + \frac{(X^5)^2}{l^2},
\end{eqnarray}
and $\kappa, \mu$ as well as $l$ are parameter{\color{fix}s} of the model. The parameter $\kappa$ characterizes the three-dimensional surface corresponding to sphere for $\kappa = +1$ for plane for $\kappa = 0$ and hyperboloid for $\kappa = -1$. The parameter $l$ characterizes the curvature of the bulk spacetime and the parameter $\mu$ characterizes the radius of a black hole in five-dimensional spacetime. Note that all parameters as well as $X^5$ are dimensionless. In order to compare this fiducial metric to the physical one, it is convenient to rescale $X^5$ to be $\phi \lambda$ where $\lambda$ is a parameter which has mass dimension. Then, the fiducial metric can be rewritten as
\begin{eqnarray}
\bar{g}_{\mu\nu} = -f(\phi) \partial_\mu \varphi^0 \partial_\nu \varphi^0 + \lambda^2 \phi^2 \Omega_{ij}(\varphi^k) \partial_\mu \varphi^i \partial_\nu \varphi^j + \frac{1}{f(\phi)} \partial_\mu \phi \partial_\nu \phi,
\end{eqnarray}
where
\begin{eqnarray}
f(\phi) = \frac{f(X^5)}{\lambda^2} = \frac{\kappa}{\lambda^2} - \frac{\mu}{\lambda^4 \phi^2} + \frac{\phi^2}{l^2}.
\end{eqnarray}
Note that $\varphi^\mu$ are St$\ddot{\text{u}}$ckelberg fields and $\phi$ is an extra physical field. It is important to note that the additional scalar degree of freedom $\phi$ is distinguish from the BD scalar since the BD scalar is always eliminated by construction of the mass terms. With general form of the fiducial metric, it was also shown that the BD ghost is not presented by using the Hamiltonian formulation \cite{Hassan:2011tf} even though the fiducial metric involves derivatives of the scalar field \cite{Andrews:2013ora}. For the Lagrangian formulation, by using the fiducial metric to be in the flat FLRW form, the scalar mode corresponding to the BD scalar is also eliminated \cite{DeFelice:2012ac}. In this work, the elimination of a scalar mode corresponding to the BD scalar will be discussed in section \ref{scalar mode section}.

In the brane point of view, this metric $\bar{g}_{\mu\nu}$ play{\color{fix}s} the role of the induced metric on the brane. By using this ansatz and unitary gauge, $\varphi^0 = t \,\,\text{and}\,\, \varphi^i = x^i$, the dRGT action in (\ref{action1}) can be expressed as
\begin{subequations}
\begin{eqnarray}
S_{dRGT} &=& 3 M_{Pl}^2 \int d^4 x \sqrt{\Omega} \, a^3 N \left(\left(\frac{\kappa}{a^2} - \left(\frac{\dot{a}}{N a}\right)^2\right) + m^2_g \left(F(X) - G(X)\frac{n}{N} \,\right) \right),\qquad \label{FLRWaction1}
\\
 S_{\text{DBI}} &=& - 3 M_{Pl}^2 m^2_g \alpha_{\Lambda} \int d^4 x \sqrt{\Omega} \, a^3 N \, X^3 \frac{n}{N}, \label{FLRWaction2}
\end{eqnarray}
\end{subequations}
where {\color{fix}$\Omega$ denotes the determinant of $\Omega_{ij}(x^k)$ and}
\begin{subequations}
\begin{eqnarray}
F(X) &=& \left(1 - X\right)\left(2 - X\right) + \frac{\alpha_3}{3}\left(1 - X\right)^2\left(4 - X\right) + \frac{\alpha_4}{3}\left(1 - X\right)^3, \\ G(X) &=& \left(1 - X\right) + \alpha_3 \left(1 - X\right)^2 + \frac{\alpha_4}{3}\left(1 - X\right)^3,
\end{eqnarray}
\end{subequations}
and we have defined
\begin{eqnarray}
X \equiv \frac{\lambda\phi}{a}, \,\,\,\,\,\, n \equiv \sqrt{f(\phi)- \frac{\dot{\phi}^2}{f(\phi)}}, \,\,\,\,\,\, \alpha_\Lambda \equiv \frac{\Lambda^4}{3 M_{Pl}^2 m^2_g}.
\end{eqnarray}

In order to obtain the equations of motion, one can vary the action in Eqs. (\ref{FLRWaction1}) and (\ref{FLRWaction2}) with respect to $N$ and $a$. By varying the action with respect to $N$, we obtain
\begin{eqnarray}
3 M_{Pl}^2\left(H^2 + \frac{\kappa}{a^2}\right) = \rho_g \equiv - 3 M_{Pl}^2 m_g^2 F.\label{eom N}
\end{eqnarray}
From this equation, one can see that the effective energy density, $\rho_g$, contributed from graviton mass is dependent on time through the function $F$ and it vanishes when $m^2_g = 0$ or $X = 1$. Varying the action with respect to $a$, the equation of motion can be written as
\begin{eqnarray}
M_{Pl}^2 \left(\frac{2 \dot{H}}{N} + 3H^2 + \frac{\kappa}{a^2} \right)= -P_g \equiv - 3M_{Pl}^2 m_g^2\left(F - \frac{X F'}{3}\left(1 - r\right)\right). \label{eom a}
\end{eqnarray}
where
\begin{eqnarray}
H \equiv \frac{\dot{a}}{a N},\,\,\,\,\,\, r \equiv \frac{n}{N X}.
\end{eqnarray}
By considering the definition of $\rho_g$ in Eq.(\ref{eom N}) and the effective pressure contributed from gravitational mass, $P_g$ in Eq.(\ref{eom a}), the solutions for accelerated expansion of the universe, $\rho_g = -P_g$, can be obtained by $F'(1-r) =0$. We will see below that $F' =0$ corresponds to the solutions in self-accelerating branch and {\color{fix}$r=1$} corresponds to a particular class of normal branch.

For the equation of motion corresponding to variation of the action with respect to the Struckelberg fields, we will use the method investigated in \cite{Gumrukcuoglu:2011zh} by expanding the action up to the linear order in $\delta \varphi^\mu \equiv \pi^\mu$ and $\delta \phi \equiv \pi^5$ without variation of the physical metric. As a result, the first order perturbation of the action can be written as
\begin{eqnarray}
\delta S = 3 M_{Pl}^2 m^2_g \int d^4 x \sqrt{\Omega} a^3 N  \Bigg[ J\, \pi^5 -  n F' \left(H - X H_\phi\right)\pi^0 \Bigg], \label{deltaS}
\end{eqnarray}
where
\begin{eqnarray}
J &\equiv& \frac{a^3}{\phi n}\Bigg{(}\frac{\left(G + \alpha_\Lambda X^3\right)\left(6f^3 + \phi f^2 f' - 3\phi \dot{\phi}^{2}f' + f\left(2\phi\ddot{\phi} - 6\dot{\phi}^{2}\right)\right)}{2 f n^2}\nonumber
\\
&&\qquad\qquad\qquad\qquad\qquad\qquad+ \left(f - N n X \left(1 + \frac{a H \phi H_\phi}{\lambda f}\right)\right)F' \Bigg{)}, \label{eom DBI field}
\\
H_\phi &\equiv& \frac{\dot{\phi}}{\phi n}.
\end{eqnarray}
Therefore, the equations of motion can be written as
\begin{eqnarray}
J &=& 0,\quad\quad\label{eom-bg3}\nonumber
\\
F' \left(H - X H_\phi\right) &=& 0. \label{eom-bg4}
\end{eqnarray}
Now we have four equations of motion and we have three variables; $a, N$ and $\phi$. However, four of them are not mutually independent due to the Bianchi identity corresponding to the conservation equation. This constraint can be written as
\begin{eqnarray}
\dot{a}\frac{\delta S}{\delta a} - N\frac{d}{dt}\left(\frac{\delta S}{\delta N}\right) +  \dot{\phi}\frac{\delta S}{\delta\phi} - \frac{d}{dt}\left(\frac{f}{n}\frac{\delta S}{\delta n}\right) = 0.
\end{eqnarray}
From Eq. (\ref{eom-bg4}), one can classify the solutions into two branches. The first  branch corresponds to $F'=0$ and the second branch corresponds to $(H - X H_\phi)=0$. For the first  branch, the equation can be expressed as
\begin{eqnarray}
F' = (-3 + 2X) - \alpha_3(1 - X)(3 - X) - \alpha_4(1 - X)^2 =  0.
\end{eqnarray}
Solving this equation, one obtains
\begin{eqnarray}
\lambda\phi = X_{\pm} a,\,\, \text{where} \,\, X_{\pm} = \frac{1 + 2\alpha_3 + \alpha_4 \pm\sqrt{1 + \alpha_3 + \alpha_3^2 - \alpha_4}}{\alpha_3 + \alpha_4},
\end{eqnarray}
which correspond to self-accelerating solutions since they provide
\begin{eqnarray}
\rho_g &=& -P_g = M_{Pl}^2 \Lambda_{\pm} \nonumber
\\
&=& -\frac{m_g^2 M_{Pl}^2}{(\alpha_3 + \alpha_4)^2}\left((1 + \alpha_3)(2+\alpha_3 + 2\alpha_3^2 -3 \alpha_4) \pm 2(1+\alpha_3 + \alpha_3^2 - \alpha_4)^{3/2} \right).
\end{eqnarray}
This branch of solutions are of the same expression with those from the dRGT massive gravity model with FLRW fiducial metric \cite{Gumrukcuoglu:2011zh}. Note that the DBI scalar field has no contribution in this self-accelerating branch which should not be so surprising since such contribution to the gravity sector must be introduced via the minimal coupling between the scalar {\color{fix}and} the massive graviton like in the mass-varying massive gravity \cite{Huang:2012pe} or the quasi-dilaton massive gravity \cite{D'Amico:2012zv}.
Note that the solutions we have so far {\color{fix}are} valid for any kind of geometry; $\kappa$ can be set to any value while the setup studied in Ref. \cite{Andrews:2013uca} and in the pure dRGT theory \cite{Gumrukcuoglu:2011ew,Gumrukcuoglu:2011zh} allows only the open slicing of the FLRW geometry.

Considering the self-accelerating branch, one can substitute the self-accelerating condition, $F'=0$, into Eq. (\ref{eom-bg3}) and then redefine variable as {\color{fix}$\dot{\phi}^2= \psi(\phi)$}. As a result, Eq. (\ref{eom-bg3}) becomes
\begin{eqnarray}
\psi' - \left(3 \frac{f'}{f}+\frac{6}{\phi}\right) \psi + f^2 \left(\frac{f'}{f}+\frac{6}{\phi}\right)=0.\label{psi eom}
\end{eqnarray}
The solution for this linear differential equation is
\begin{eqnarray}
\dot{\phi}^2 = \psi = f^2 -\left(\frac{\phi}{\phi_0}\right)^6 f^3,\label{phidot sol}
\end{eqnarray}
where $\phi_0$ is an integration constant. From this solution, we can find the relation of the lapse function in terms of the DBI scalar field by using Eq. (\ref{eom N}) as follows
\begin{eqnarray}
N^2 = \frac{3\lambda^{2}}{\left(\lambda^{2}\phi^{2}\Lambda_{\pm} - 3\kappa X_{\pm}^{2}\right)}\left(f^2 - \left(\frac{\phi}{\phi_0}\right)^6 f^3\right).\label{N sol}
\end{eqnarray}
Moreover, we can investigate the possible interval of the DBI scalar by considering $\dot{\phi}^2 > 0$. Its interval is determined by $\lambda$, $\mu$, $l$, $\kappa$ and $\phi_0$. In case of the flat universe, it can be written in the {\color{fix}simple} form {\color{fix}as}
\begin{eqnarray}
0 < \phi < \frac{b^{1/2}}{2^{1/4} }\left(1 + \sqrt{1 + \frac{4 \lambda^4 \phi_0^6}{\mu}}\right)^{1/4},
\end{eqnarray}
where $ b = \sqrt{\mu l^2 / \lambda^4}$. From Eq. (\ref{phidot sol}), one can see that the fixed points of the system are $f=0$ and $f = (\phi_0 / \phi)^6$. The scalar field will evolve to the points $\phi^2 = 0$ or $\phi^2 = b$ for the fixed point $f=0$. For the fixed point  $f = (\phi_0 / \phi)^6$, the scalar field will evolve to the points $\phi^2 = \frac{b}{\sqrt{2} }\left(1 \pm \sqrt{1 + (4 \lambda^4 \phi_0^6)/\mu }\right)^{1/2}$.  For the self-accelerating branch, it seems like that the dynamics of the universe is not controlled by the property of the fiducial metric since the Hubble parameter $H$ is always constant and does not depend on function $f$. In other words, no matter how the scalar field $\phi$ evolves, the Hubble parameter $H$ is always constant.

For the second branch, namely normal branch, the equations of motion are very complicated. However, we can restrict our attention to a particular class of the solutions which simplifies the calculation while still provide the significant result of the system. The characteristic equation of this branch is $H - X H_\phi =0$,  so that we obtain $X = H/H_\phi$. From Eq. (\ref{eom-bg3}), we choose the solution such that $G + \alpha_\Lambda X^3 = 0$ to simplify the calculation as well as to capture the significant dynamics of the universe. This equation also satisfies the equation obtained by varying the action with respect to $n$, $\delta S /\delta n = 0$, and can be expressed in terms of $X$ as
\begin{eqnarray}
 \left(\alpha _4-3 \alpha _{\Lambda }\right)X^3 -3 \left(\alpha _3+\alpha _4\right) X^2 +3\left(1+2 \alpha _3+\alpha _4\right) X-(3 +3 \alpha _3+\alpha _4)=0.\label{equ-nor}
\end{eqnarray}
Solving this equation, one obtains the relation between $\phi $ and $a$ as
\begin{eqnarray}
\lambda \phi  = c_3 a, \label{sol-nor1}
\end{eqnarray}
where $c_3$ is a constant depending on $\alpha_3, \alpha_4$ and $\alpha_\Lambda$. Since $G + \alpha_\Lambda X^3 = 0$ is the third order equation, $c_3$ can take three values to satisfy this equation. There is a particular value of $\alpha_\Lambda$ such that $\alpha _4 = 3 \alpha _{\Lambda }$ which provides only two solutions of $X$. Substituting this relation into the equation $H - X H_\phi =0$, one obtains
\begin{eqnarray}
\frac{n}{N} = c_3.\label{sol-nor2}
\end{eqnarray}
This solution corresponds to $r = 1$. Note that all solutions where $X$ is constant in normal branch  provide the constraint of $r = 1$. As we have mentioned before, from Eqs. (\ref{eom N}) and (\ref{eom a}), one can see that the constraint of $r = 1$ provides the relation $P_g = -\rho_g = \text{constant}$, corresponding to an accelerated expansion of the universe. This solution provides the same behavior like the solutions in self-accelerating branch. However, the solutions are different since the constant $c_3$ is not generally equal to $X_{\pm}$. Moreover, the dynamics of the scalar field $\phi$ in both branches are different. In order to find the dynamics of the scalar field in normal branch for this solution, one can substitute relations in Eq. (\ref{sol-nor1}) and Eq. (\ref{sol-nor2}) into Eq. (\ref{eom N}). As a result, the equation for the scalar field can be written as
\begin{eqnarray}
\dot{\phi}^2 = \frac{c_n \phi^2 f^2}{1+c_n\phi^2},\label{eq-nor-phi}
\end{eqnarray}
where
\begin{eqnarray}
c_n = \frac{\rho_g }{3 M_{Pl}^2 c_3^2}.
\end{eqnarray}
Note that we considered this solution in the flat geometry where $\kappa = 0$ for simplicity. This equation can be viewed as autonomous equation which has a fixed point at $f = 0$ corresponding to the point $\phi^2 = b$. To obtain the full behavior of the scalar field, one can solve Eq. (\ref{eq-nor-phi}) and the solution can be expressed as
\begin{align}
\sqrt{1+c_n b } \,\tanh^{-1}\left( \sqrt{\frac{1 + c_n \phi^2 }{1+c_n b}}\right) - \sqrt{1-c_n b } \,\tanh^{-1}&\left( \sqrt{\frac{1 + c_n \phi^2 }{1-c_n b}}\right) \nonumber
\\
&= \frac{2 b \sqrt{c_n} }{l^2 } t + C,\label{sol-nor-phi}
\end{align}
where $C$ is an integration constant determining the initial value of the field when initial time is taken. From this expression, we found that $\phi^2 \rightarrow b$ as $t \rightarrow \infty$. This is what we expect from the analysis in {\color{fix}the} autonomous system. It is important to note that our analysis for normal branch is only a particular class of the solutions in the branch. Other solutions which may provide the result such that $X = X(t)$ is much more complicated than this solution and thus is not easy to study analytically. They may give some interesting results and we leave this issue for further work.

\section{Perturbations\label{sec:pert}}
To investigate the stability of the model, we expand the action perturbatively up to the quadratic order.
Here, the physical metric around which the action is expanded quadratically is expressed as follow,
\begin{eqnarray}
g_{\mu\nu}=g_{\mu\nu}^{(0)}+ \delta g_{\mu\nu},
\end{eqnarray}
where the $g_{\mu\nu}^{(0)}$ is a flat FLRW metric and the perturbations are decomposed as
\begin{subequations}
\begin{eqnarray}
\delta g_{00} &=& -2N^2 \Phi,\\ \delta g_{0i} &=& N\, a (B^T_i + \partial_i B),\\ \delta g_{ij} &=& a^2 \left[ h_{ij}^{TT} + \frac{1}{2} (\partial_i E^T_j + \partial_j E^T_i) + 2\delta_{ij} \Psi + \Big( \partial_i \partial_j - \frac{1}{3} \delta_{ij} \partial_k\partial^k \Big) E \right].
\end{eqnarray}
\end{subequations}
where $\Phi, B, \Psi, E$ are scalar parts, $B_i^T, E_i^T$ are transverse vector parts, and $h_{ij}^{TT}$ is a transverse-traceless tensor part.
We choose the unitary gauge in our analysis which corresponds to setting $\pi^\mu = 0$. Thus the other perturbation comes from the DBI scalar field, $\delta \phi$. For simplicity, we choose to work in flat FLRW universe. In the following we will investigate the stabilities of the perturbations in tensor, vector and scalar modes separately.

\subsection{Tensor modes}
\begin{figure}[h!]
\centering
\begin{subfigure}[b]{0.35\textwidth}
\centering
\includegraphics[width=1.\textwidth]{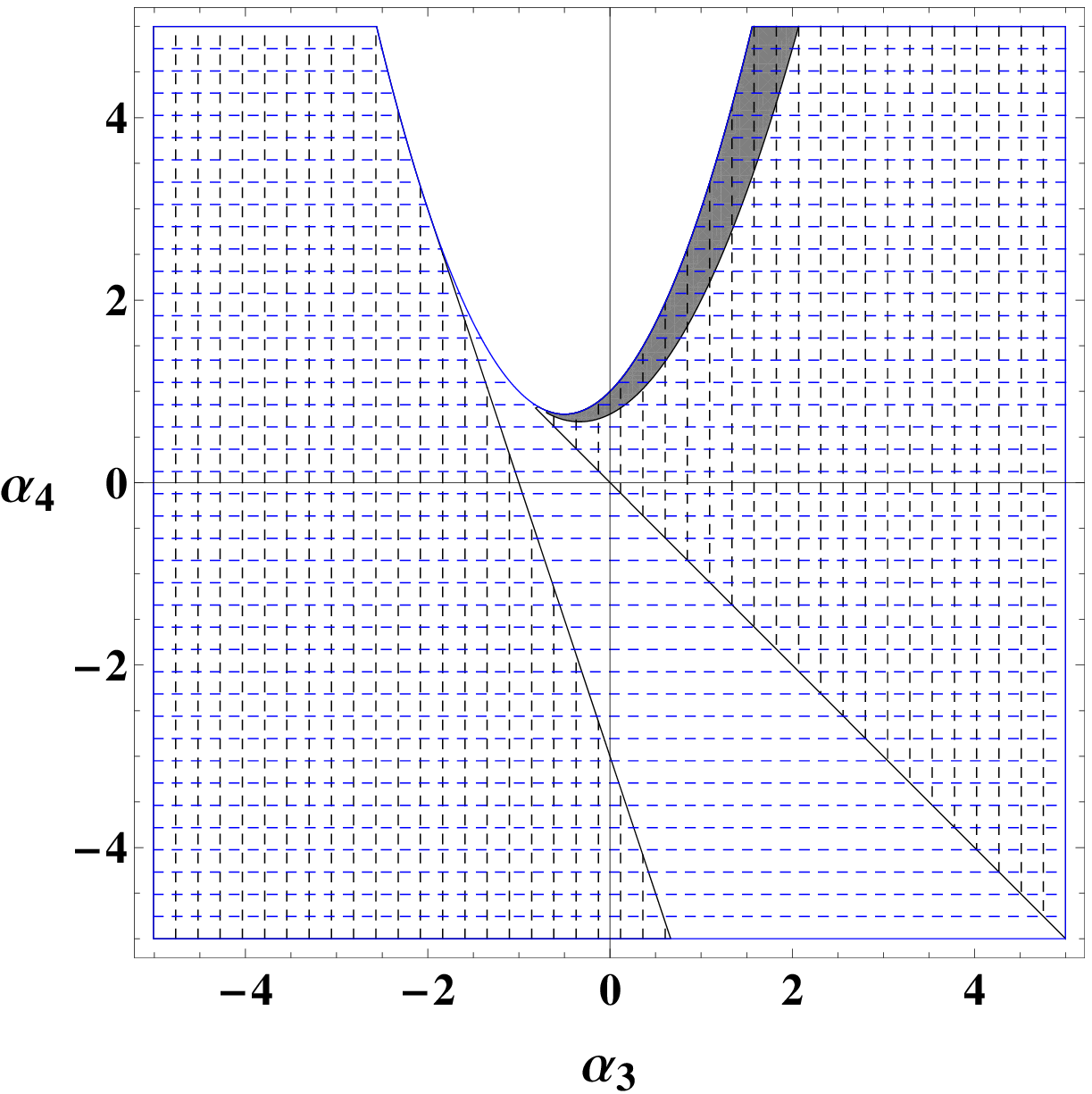}
\caption{$X_{+}$ solution}
\label{fig:self-Acc-p}
\end{subfigure}
\begin{subfigure}[b]{0.35\textwidth}
\centering
\includegraphics[width=1.\textwidth]{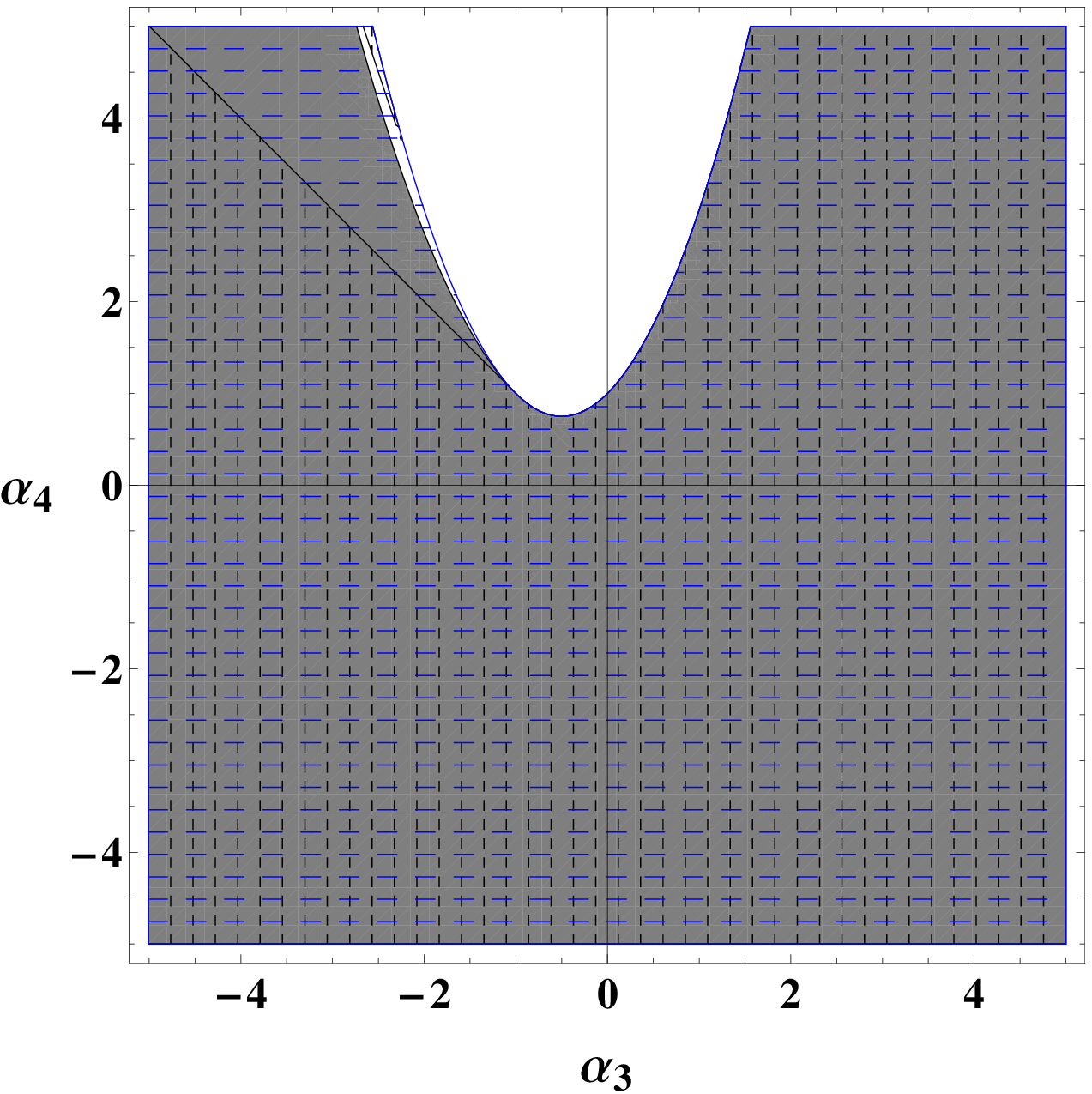}
\caption{$X_{-}$ solution}
\label{fig:self-Acc-m}
\end{subfigure}
\caption{Allowed regions in $(\alpha_3, \alpha_4)$ space for  {\color{fix}the self-accelerating} branch plotted by using three conditions: $M^2_{GW} > 0$, $\rho_g > 0$ and $X_{\pm} > 0$. The left panel corresponds to $X_{+}$ solution and the right panel corresponds to $X_{-}$ solution. The shaded region with horizontal-dashed-blue line correspond to the condition $M^2_{GW} > 0$. The shaded region with vertical-dashed-black line corresponds to the condition $X_{\pm} > 0$. The grey region corresponds to the condition $\rho_g > 0$.}
\label{fig:self-Acc}
\end{figure}

For the tensor modes, we expand action in Eq. (\ref{action1}) up to second order of $h_{ij}^{TT}$. After that we keep only the second order and then transform the perturbation variables to {\color{fix}those in} the Fourier space. As a result, the second order of the action for the tensor modes in Fourier space can be written as
\begin{eqnarray}
S^{(2)} = \frac{M_{Pl}^2}{8}\int d^3k \,dt \,a^3\, N\, \left(\frac{|\dot{h}_{ij}^{TT}|^2}{N^2} -\Big(\frac{k^2}{a^2} + M^2_{GW}\Big)|h_{ij}^{TT}|^2 \right),
\end{eqnarray}
where
\begin{subequations}
\begin{eqnarray}
M^2_{GW} &=& m^2_g A,
\\
A &=& X^2 \Bigg{(}\left(\frac{3}{X} - 1 - r\right) + \alpha_3\left(\left(\frac{3}{X} - 2\right) - \left(2 - X\right)r\right) \nonumber
\\
&&\qquad+ \alpha_4\left(1 - X\right)\left(\frac{1}{X} - r\right)\Bigg{)}.
\end{eqnarray}
\end{subequations}
The kinetic term of the tensor mode always has the correct sign. Thus there is no ghost in this mode. The absent of tachyonic instability is required by the condition $M^2_{GW} = m^2_g A > 0$. For the self-accelerating branch, this condition depends only on $\alpha_3$ and $\alpha_4$. We use this condition together with $\rho_g > 0$ and $X_{\pm} > 0$ in order to find compatible regions in the ($\alpha_3,\alpha_4$) space. The allowed regions for both solutions are shown in Fig. \ref{fig:self-Acc}. The left panel corresponds to $X_{+}$ solution and the right panel corresponds to $X_{-}$ solution. The shaded region with horizontal-dashed-blue line correspond to the condition $M^2_{GW} > 0$. The shaded region with vertical-dashed-black line corresponds to the condition $X_{\pm} > 0$. The grey region corresponds to the condition $\rho_g > 0$. For the condition $M^2_{GW} > 0$, it is found that the region will exist if $r > 1$ for $X_{+}$ solution and $r < 1$ for $X_{-}$ solution. For $r = 1$, it corresponds to $M^2_{GW} = 0$ which is reduced to massless gravity theory. Since the solutions in this branch are not governed by the DBI-scalar field, the region are the same with dRGT massive gravity with FLRW fiducial metric \cite{Gumrukcuoglu:2011zh}.

\begin{figure}[h!]
\centering
\begin{subfigure}[b]{0.35\textwidth}
\centering
\includegraphics[width=1.\textwidth]{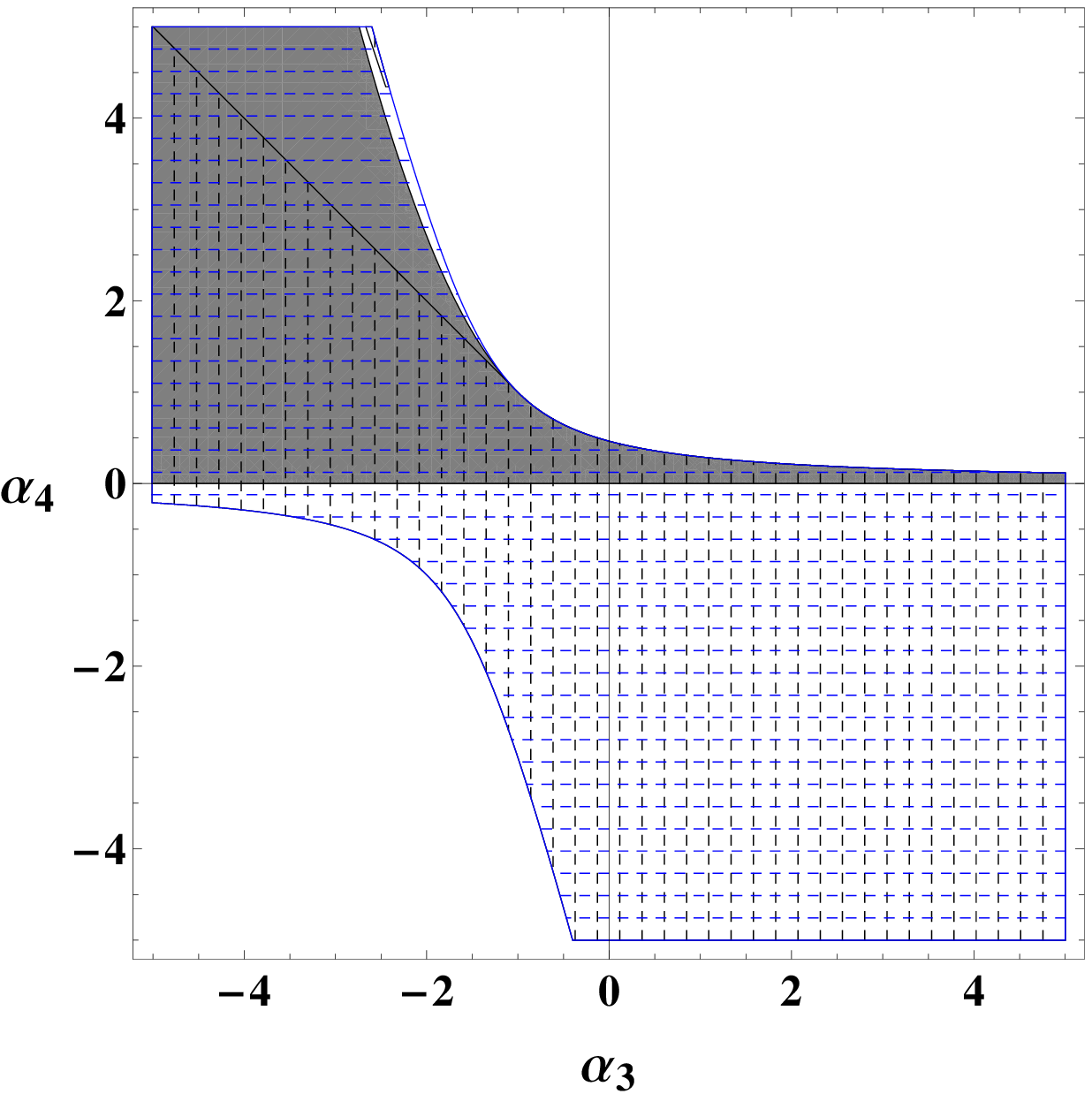}
\caption{$X_{-}$ solution with condition $M^2_{GW} > 0$}
\label{normal-m1}
\end{subfigure}
\begin{subfigure}[b]{0.35\textwidth}
\centering
\includegraphics[width=1.\textwidth]{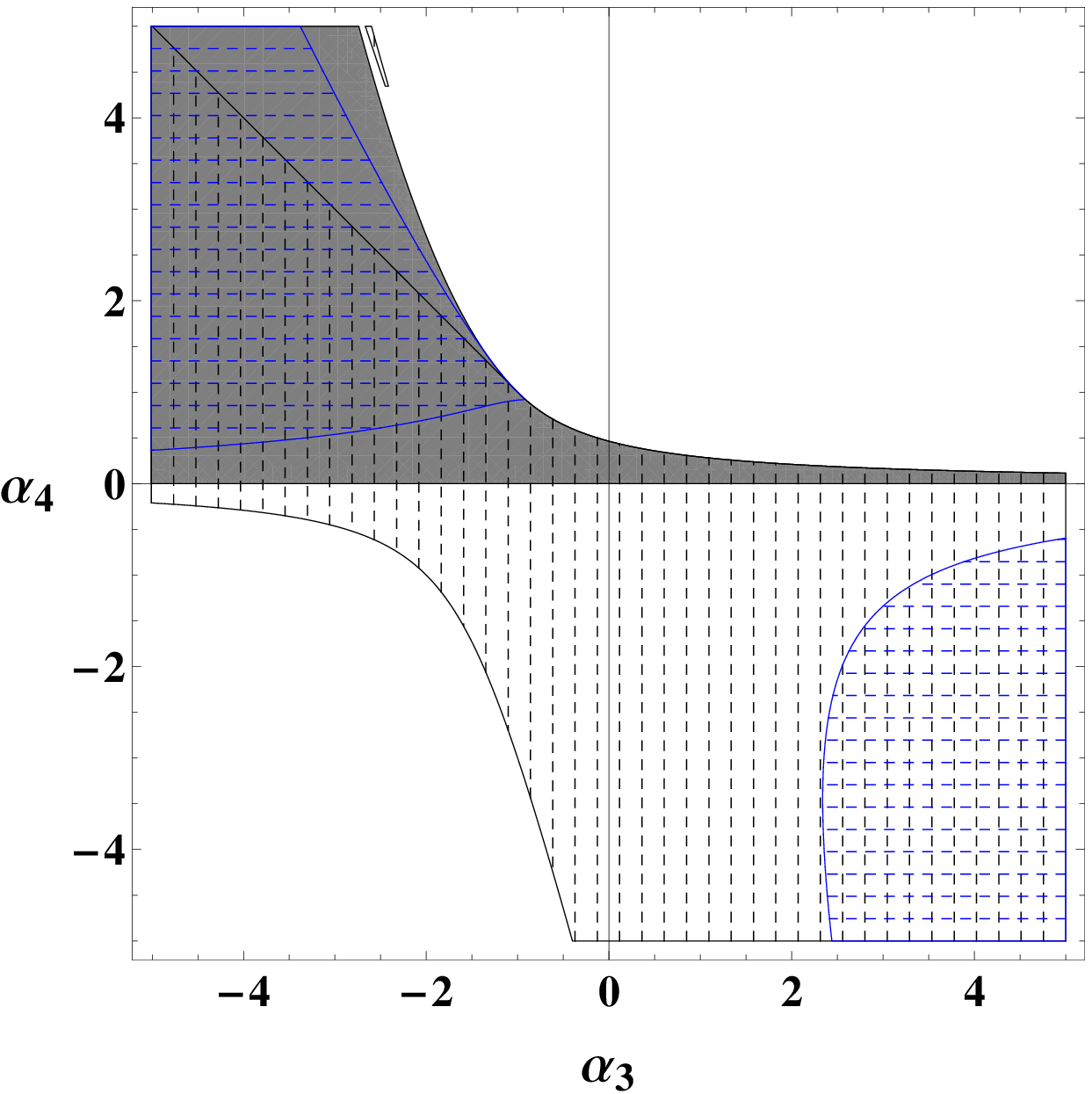}
\caption{$X_{-}$ solution with condition $\frac{M^2_{GW}}{ H^2 } > 2 $}
\label{normal-m2}
\end{subfigure}
\caption{Allowed region in $(\alpha_3, \alpha_4)$ space for normal branch plotted by using three conditions,  $M^2_{GW} > 0$, $\rho_g > 0$ and $X_{-} > 0$ for the left panel and $M^2_{GW}/H^2 > 2$, $\rho_g > 0$ and $X_{-} > 0$ for the right panel. The shaded region with horizontal-dashed-blue line correspond to the condition $M^2_{GW} > 0$ in the left panel and $M^2_{GW}/H^2 > 2$ for the right panel. The shaded region with vertical-dashed-black line corresponds to the condition $X_{\pm} > 0$. The grey region corresponds to the condition $\rho_g > 0$.}
\label{fig:normal}
\end{figure}

For the normal branch, we choose the particular class of the solutions which discuss in the previous section. These solutions depend on three parameters, $\alpha_3$, $\alpha_4$ and $\alpha_\Lambda$. For simplicity, we also restrict our attention only in the case $\alpha_\Lambda = \alpha_4 /3$. This restriction provides two solutions by solving Eq. (\ref{equ-nor}). As a result, the allowed region which satisfies the condition $M^2_{GW} > 0$, $\rho_g > 0$ and $X_{-} > 0$ is shown in Fig. \ref{normal-m1}. Note that there is no allowed region to satisfy the condition $\rho_g > 0$ for the solution $X_{+} $. Surprisingly, the graviton mass for these solutions coincides with the effective mass investigated by using de Sitter and FLRW fiducial metric in dRGT massive gravity \cite{Fasiello:2012rw}. This effective mass is constrained by Higuchi bound as $M_{GW}^2 > 2 H^2$. We use this condition together with $\rho_g > 0$ and $X_{-} > 0$ to find the allowed region. The result is shown in Fig \ref{normal-m2}. At this point, one can see that we can find the viable model by specifying the parameters $\alpha_3$ and $\alpha_4$ in both branches. However, by investigating the perturbation in vector and scalar mode, it is found that the theory is suffered from ghost instability. We explore this behavior in the next two subsections.

\subsection{Vector modes}
For vector modes, we follow the calculation step as performed in tensor modes. As a result, the second order action in Fourier space can be written as
\begin{align}
S^{(2)} &= \frac{M_{Pl}^2}{2}\int d^3k  \,dt \,a^3\, N\, \Bigg\{  \frac{k^2}{8 N^2} |\dot{E}^T_i|^2 - \frac{k^2}{2 N a} B^T_i \dot{E}^T_i - \frac{k^2}{4}\left(\frac{2 \dot{H}}{N} + 3H^2\right)|E^T_i|^2 \nonumber
\\
&\qquad+ \left(\frac{k^2}{2 a^2} + 3H^2\right)|B^T_i|^2 - \frac{k^2}{4}\left(\frac{P_g}{M_{Pl}^2} + \frac{1}{2} M_{GW}^2\right)|E^T_i|^2 - \left(\frac{\rho_g}{M_{Pl}^2} + \frac{m_g^2 X F'}{(1 + r)}\right)|B^T_i|^2 \Bigg\} \label{actionO2-vector-EH}.
\end{align}
From this action, one can see that $B^T_i$ is non-dynamical. Thus we can algebraically solve it and then substitute its solution back to the action. As a result, the action becomes
\begin{eqnarray}
S^{(2)} = \frac{M_{Pl}^2}{8}\int d^3k \,dt \,a^3\, N\, \Bigg\{ -\frac{k^2 m^2_g a^2 X F'}{N^2\left(k^2(1 + r) - 2 m^2_g a^2 X F'\right)} |\dot{E}^T_i|^2 - \frac{k^2}{2} M_{GW}^2 |E^T_i|^2 \Bigg\}.\qquad \label{actionO2-vector}
\end{eqnarray}
It is found that, for self-accelerating branch; $F'=0$, there is no propagating d.o.f in vector modes. For the normal branch, it is ghost free if $m^2_g F' < 0 $. In the case of $m^2_g F' > 0 $, the propagating vector seems to suffer from ghost at small scale or large $k$ when
\begin{eqnarray}
k^2 \geq \frac{2 a^2 X F'}{(1 + r)}m^2_g \sim m_g^2.
\end{eqnarray}
The sound speed of propagation can be written as
\begin{eqnarray}
c^2_V = -\frac{M^2_{GW}(1 + r)}{2 m^2_g X F'} = - \frac{A(1 + r)}{2 X F'}.
\end{eqnarray}
Thus, the condition to avoid gradient instability is
\begin{eqnarray}
 \frac{A}{F'} < 0.
\end{eqnarray}
It can be summarized that the condition for avoiding the instabilities is that $ m^2_g A > 0,  \frac{A}{F'} < 0$ or corresponding to $m^2_g F' < 0$.

\subsection{Scalar modes}\label{scalar mode section}
For scalar modes, the second order perturbations in Fourier space for Einstein-Hilbert can be written as
\begin{subequations}
\begin{align}
S^{(2)}_{EH} &= \frac{M_{Pl}^2}{2} \int d^3k \,dt \,a^3\, N\, \Bigg\{ \frac{k^4 \dot{E}^2}{6 N^2} - \frac{6 \dot{\Psi}^2}{N^2} - \frac{2 k^4 B\dot{E}}{3 a N} + \dot{\Psi}\left(\frac{12 H \Phi }{N} - \frac{4 k^2 B}{a N}\right) \nonumber
\\
&\qquad+ E^2\left(\frac{k^6}{18 a^2} - \frac{k^4}{3}\left(\frac{2 \dot{H}}{N} + 3 H^2\right)\right) + \Psi^2\left(\frac{2 k^2}{a^2} + 3\left(\frac{2\dot{H}}{N} + 3 H^2\right)\right) \nonumber
\\
&\qquad+ 3 k^2 H^2 B^2 - 9 H^2 \Phi^2 + \frac{2 k^4 \Psi E}{3 a^2}+ \Phi\left(\frac{4 k^2 H B}{a} + \Psi\left(\frac{4 k^2}{a^2} + 18 H^2\right) + \frac{2 k^4 E}{3 a^2}\right) \Bigg\} \label{actionO2-EH}.
\end{align}
For dRGT and DBI parts, the action can be expressed as
\begin{align}
S^{(2)}_{dRGT} + S^{(2)}_{DBI} &= \frac{M_{Pl}^2 m^2_g}{2} \int d^3k \,dt \,a^3\, N\, \Bigg\{3\left(\frac{G + \alpha_\Lambda X^3}{N n^{3}}\right)\dot{\delta\phi}^2 - \frac{6 \phi H_\phi F'}{N f}\dot{\delta\phi}\Psi - \frac{3 J_{,\phi}}{a^3 N}\,\delta\phi^2 \nonumber
\\
&\quad- \frac{k^2}{a^2 N n X}\left(\left(G' + 3 \alpha_\Lambda X^2\right) - \frac{(f + N n X) F'}{f X (1 + r)}\right)\delta\phi^2 - \frac{k^4}{3}\left(\frac{P_g}{M_{Pl}^2 m_{g}^2} + \frac{A}{2}\right)E^2 \nonumber
\\
&\quad+ 3\left(\frac{P_g}{M_{Pl}^2 m_{g}^2} + 2 A\right)\Psi^2 + k^2\left(\frac{P_g}{M_{Pl}^2 m_{g}^2} - \frac{n^2 F'}{N^2 X (1 + r)}\right)B^2 + \frac{\rho_g}{M_{Pl}^2 m_{g}^2}\Phi^2 \nonumber
\\
&\quad+ 3\left(\frac{f'(f^2 + \dot{\phi}^2)F'}{N n f^2} - \frac{4 A}{\phi}\right)\delta\phi\Psi + \frac{2 k^2 n H_\phi F'}{N \lambda f (1 + r)}\delta\phi B + \frac{6 \lambda F'}{a}\,\delta\phi\Phi \nonumber
\\
&\quad+ 18\left(F - \frac{X F'}{3}\right)\Phi\Psi \Bigg\}. \label{actionO2-dRGT}
\end{align}
\end{subequations}

From the action in (\ref{actionO2-EH}) and (\ref{actionO2-dRGT}), $B$ and $\Phi$ are non-dynamical. We can use their equations of motion to eliminate them from the action. These constraints comes from the energy and scalar part of the momentum conservation. Now we have three variables $\delta \phi, E$ and $\Psi$. One of them is a scalar degree of freedom for the massive graviton, the others are BD scalar and DBI scalar. However, we found that the kinetic term of $\Psi$ will vanish by imposing the background equation (\ref{eom N}) and (\ref{eom a}). Therefore, the scalar  $\Psi$ can be interpreted as the BD scalar since it must be eliminated by construction of the theory. We can eliminate it from the action by following the previous procedures. Note that the full lagrangian before integrating out $\Psi$ is quite lengthy and then we left in the Appendix \ref{fullaction}, Eq.(\ref{actionO2-scalar}). Generally, we have the scalar mode action of the form
\begin{eqnarray}
S &=& \int d^3k dt \left(K_{IJ}\dot{\chi}^I\dot{\chi}^J + M_{IJ}\dot{\chi}^I\chi^J + P_{IJ}\chi^I\chi^J\right),
\end{eqnarray}
where $\chi_1 = \delta\phi$ and $\chi_2 = E$. The $K_{IJ}, M_{IJ}$ and $P_{IJ}$ each are elements of corresponding $2\times 2$ matrices associated to each combinations of the remaining propagating d.o.f.

For the self-accelerating branch, we have $F' = 0 = \dot{H}$. It turns out that according to this branch, there is no propagating scalar d.o.f in gravity sector; only propagating scalar is obviously the DBI scalar. The scalar mode action for the self-accelerating branch is specified with the matrix $K_{IJ}$ as follows,
\begin{eqnarray}
K^{(s)}_{11}=\frac{3M^2_{Pl}m^2_g a^3 \left(G+\alpha_\Lambda X^3\right)}{2 n^3},\qquad K^{(s)}_{12}=
K^{(s)}_{22}=0,\qquad \det{(K^{(s)}_{IJ})}=0.
\end{eqnarray}
which the only nonvanishing matrix element is $K_{11}$ corresponding to the DBI scalar $\delta \phi$ as claimed.

For the normal branch, we consider the case which satisfies the condition we considered previously,
\begin{eqnarray}
G + \alpha_\Lambda X^3 = 0,\quad r=1.
\end{eqnarray}
Even though this equation is not independent to the other equations of motion, we still use this equation in order to simplify the second order action. By imposing this background equation of motion, it seems like that the DBI-scalar d.o.f does not propagate. However, substituting back the solution of $\Psi$, one obtains
\begin{eqnarray}
K^{(n)}_{11} &=& - \frac{9 M^2_{Pl} m^2_g a \lambda^2 V^2 F'^2}{N H^2 W}, \nonumber
\\
K^{(n)}_{12} &=& \frac{3 M^2_{Pl} k^4 \lambda  V F'^2}{N H^2 U W}, \nonumber
\\
K^{(n)}_{22} &=& M^2_{Pl}\left(\frac{k^4 a^3 F'}{4 N U} - \frac{k^8 F'^2}{m^2_g a N H^2 U^2 W}\right), \nonumber
\\
\det{(K^{(n)}_{IJ})} &=& - \frac{9 M^4_{Pl} m^2_g k^4 a^4 \lambda^2 V^2 F'^3}{4 N^2 H^2 U W},
\end{eqnarray}
where
\begin{eqnarray}
U &=& \left(3 F' - \frac{4 k^2}{m^2_g a^2 X}\right), \,\,\,\,\,\, V = \left(1 + \frac{a^2 H^2 }{\lambda^2 f}\right), \nonumber \\ W &=& \Bigg(12 A + 2\left(\frac{2 k^4}{m^2_g a^4 H^2 U} + 9 X\right)F' + \frac{3 m^2_g X^2 F'^2}{H^2} \Bigg).
\end{eqnarray}
It is found that, for the particular class of the solutions in normal branch, the d.o.f of the theory is still correct. At small scale, $U \propto -1/m^2_g$  and then $W \propto -F'$ where the proportional constant is positive. Substituting these results into $K^{(n)}_{11}$, it is found that $K^{(n)}_{11} \propto m^2_g F'$ where the proportional constant is positive. This leads to the condition to avoid ghost such that $ m^2_g F' > 0$. This condition is in conflict with the condition from the vector mode in which $ m^2_g F' < 0$. As a result, the ghost degree of freedom always exist in this branch at small scale (either in vector modes or in one of scalar modes). For the other mode of scalar perturbations, one can find the result by substituting $U$ and $W$ at small scale into $\det{(K^{(n)}_{IJ})}$ . As a result, we found that $\det{(K^{(n)}_{IJ})}$ is always negative implying that, at least, one of scalar mode is always ghost. It is important to note that the leading term of $K^{(n)}_{11}$ is proportional to $k^{-2}$ leading to $K^{(n)}_{11} \rightarrow 0$ when we take $k \rightarrow \infty$. This behavior occurs only in the particular solutions we choose to consider since the leading term, corresponding to $k^{0}$, is proportional to $G + \alpha_\Lambda X^3$. This suggests that the contributions from linear perturbations are very small. Therefore non-linear perturbations may have nontrivial contributions and the conclusion of our results may change when non-linear perturbations are taken into account.

\section{Conclusion\label{sec:con}}

In this work, we have considered the extended DBI massive gravity by generalizing the fiducial metric. In the original version of DBI massive gravity, the fiducial metric is in Minkowski form. It was found that the flat and closed solutions for FLRW metric {\color{fix}are not admitted}. Moreover, the number of degrees of freedom of the massive graviton {\color{fix}is} not correct. The flat Minkowski fiducial metric in DBI massive gravity plays the role of the induced metric on the brane. In the present work, we generalize the fiducial metric by considering the induced metric corresponding to a domain wall moving in five-dimensional Schwarzschild-AdS spacetime.

From background equations, we found that the solutions not only admit open geometry of FLRW metric but also flat and closed geometry. Moreover, the solutions can be divided into two branches; self-accelerating branch and normal branch. For self-accelerating branch, the graviton mass {\color{fix}plays} the role of cosmological constant to drive the late-time acceleration of the universe. We found that the Hubble parameter $H$ is always constant and does not depend on function $f$ which characterizes the form of the fiducial metric. Therefore, the dynamics of the universe are not controlled by the property of the fiducial metric. Furthermore, the DBI scalar field has no contribution to the solutions in this branch. This is not so surprising since such contribution to the gravity sector must be introduced by the minimal coupling between the scalar and the massive graviton like in the mass-varying massive gravity and quasi-dilaton massive gravity.  In order to investigate the number of degrees of  freedom of the theory, we performed cosmological perturbations around this background solutions. From the tensor modes, we have explored the conditions to avoid the tachyon{\color{fix}ic} instability, $M_{GW}^2 >0$, while the ghost and gradient instabilities are absent. By using this condition together with consistent conditions, $\rho_g > 0$ and $X_\pm > 0$, we found that there are allowed regions in $(\alpha_3, \alpha_4)$ space in both $X_{+}$ and $X_{-}$ solutions as shown in Fig. \ref{fig:self-Acc}. By including the investigation of vector and scalar modes, we found that the number of degrees of of freedom is not correct. There are only two degrees of freedom for gravity sector contributed from tensor modes while the contribution from vector and scalar modes vanish. This inconsistency is similar to the dRGT massive gravity and DBI massive gravity.

For the solutions in normal branch, the equations of motion are complicated and it is not easy to obtain the analytical solutions. Therefore, the solutions in this branch are quite lack of investigation in literature. This is not only due to complicated equations but also the solutions generally do not provide an accelerated expansion of the universe. In the present work, we restrict our attention to a particular class of the solution in order to simplify the calculation and also to capture some significant behaviors of the solutions. A characteristic constraint of this class of the solutions is $r = n/(NX) = 1$. We found that this class of the solutions also provides the accelerat{\color{fix}ing} expansion of the universe like in self-accelerating branch inferred from the definition of effective energy density $\rho_g$ and the effective pressure $P_g$ in Eqs. (\ref{eom N}) and (\ref{eom a}) respectively. Even though the solutions in this class provide the same accelerated expansion of the universe like the solutions in self-accelerating branch, these solutions are different since the solutions in this branch depend on the effect of DBI scalar field via the coupling $\alpha_\Lambda$. By setting the coupling $\alpha_\Lambda = \alpha_4 /3$ to simplify the solutions, the allowed regions plotted by using conditions $\rho_g > 0$, $X_{-} > 0$ and  $M_{GW}^2 >0$ are explored. Moreover, we also found that the graviton mass in this branch coincides with the effective mass investigated by using de Sitter and FLRW fiducial metric in dRGT massive gravity \cite{Fasiello:2012rw}. This effective mass is constrained by Higuchi bound as $M_{GW}^2 > 2 H^2$. We use this condition together with $\rho_g > 0$ and $X_{-} > 0$ to find the allowed region in $(\alpha_3, \alpha_4)$ space shown in Fig \ref{fig:normal}. By using the investigation of the perturbation in vector and scalar modes, we found that the number of degrees of freedom is correct. However, at small scale with $k^2 \gtrsim m^2_g$, the conditions to avoid ghost instability from vector modes and scalar modes are {\color{fix}in} conflict implying that there exists at least one ghost degree of freedom. This investigation suggests that the extended model of DBI massive gravity in this way dose not provide a viable cosmological model to explain the late-time acceleration of the universe. In order to obtain a viable extended model of DBI massive gravity, one may introduce the coupling term similar to one in the generalized quasidilaton massive gravity or introduce the nontrivial coupling matter fields like in original dRGT massive gravity.

\begin{acknowledgments}
The authors are grateful to Shinji Mukohyama and Antonio De Felice for an initiation of this work, helpful conversations and comments regarding to the manuscript. Pitayuth Wongjun and Lunchakorn Tannukij are supported by Thailand Toray Science Foundation (TTSF) from science and technology research grant.
\end{acknowledgments}

\appendix

\section{Full quadratic scalar mode action\label{fullaction}}
The full quadratic scalar mode action, after imposing the background equations of motion (\ref{eom N}, \ref{eom a}), is
\begin{align}
\displaystyle
S^{(2)} &= M_{Pl}^2 \int d^3k \, dt a^3 N \Bigg\{ \frac{3m^2_g}{2}\left(\frac{G + \alpha_\Lambda X^3}{N n^3}\right)\dot{\delta\phi}^2 +\frac{k^4 F'}{4 N^2 U}\dot{E}^2 + \frac{k^4 r F'}{a^2 N H U}\Psi\dot{E}  \nonumber
\\
&\quad - \frac{3 m^2_g \lambda V F'}{a N H}\dot{\delta\phi}\Psi - \frac{k^4 (\lambda^2 f(1 + r) + a^2 r X H H_\phi)F'}{a^3 N \lambda f X H U}\delta\phi\dot{E}  \nonumber
\\
&\quad+ \frac{k^4}{18}\left(\frac{k^2}{2 a^2} - \frac{M^2_{GW}}{4} + \frac{k^4 r^2 F'}{2 a^4 H^2 U} - \frac{k^4 \left((1 + r)\phi\dot{U} + \left((1 + r)\dot{\phi} - \phi\dot{r}\right)U\right)}{m^2_g a^4 N \phi X H U^2}\right)E^2 \nonumber
\\
&\quad- \frac{m^2_g}{2}\Bigg[\frac{k^2(G' + 3 \alpha_\Lambda X^2)}{a^2 N^2 r X^2} + \frac{3 J_{,\phi}}{a^3 N} - \frac{F'}{2 a^2 N^2}\Bigg(\frac{2 k^2(f + N^2 r X^2)}{f r(1 + r) X^3} + \frac{9 m^2_g N^2 \lambda^2 F'^2}{H^2 U} \nonumber
\\
&\quad+ \frac{6 k^2 N^2 \phi r H\left(2\lambda^2 f(1 + r) + a^2 r X H H_\phi\right)F'}{\lambda^2 f^2 H(1 + r)U} \Bigg)\Bigg]\delta\phi^2 + \frac{m^2_g W}{4}\Psi^2\nonumber
\\
&\quad  + \frac{k^4(2 k^2 r H H_\phi + 3 m^2_g \lambda^2 f F')F'}{6 a^3 \lambda f H^2 U}\delta\phi E - \frac{k^4 r(2 k^2 + m^2_g a^2 X U)F'}{6 a^4 H^2 U}\Psi E \nonumber
\\
&\quad- \Bigg[6\frac{M^2_{GW}}{\phi} + \left(\frac{6 m^2_g \lambda}{a} - \frac{2 k^4 \phi r H}{a^3 \lambda f H U} - \frac{3 m^2_g f'(f^2 + N^2 \phi^2 X^2 r^2 H_\phi^2)}{2 N^2 f^2 r X}\right)F' \nonumber
\\
&\quad- \frac{3 m^2_g \lambda X}{a}\left(1 - \frac{X r H_\phi}{H}\right)F'' - \frac{3 m^2_g \left(\lambda^2 f(\frac{2 k^2}{a^2} + m^2_g(1 - r)U) - 2 k^2 X r H H_\phi\right)F'^2}{2 a \lambda f H^2 U}\nonumber
\\
&\quad + \frac{9 m^4_g \lambda X F'^3}{2 a H^2 U}\Bigg]\delta\phi\Psi\Bigg\}.  \label{actionO2-scalar}
\end{align}
By solving $\Psi$ and then substituting back into the action, the components of the kinetic terms matrix can be written as
\begin{eqnarray}
K_{11} &=& - \frac{9 M^2_{Pl} m^2_g a \lambda^2 V^2 F'^2}{N H^2 W} +\frac{3M^2_{Pl}m^2_g a^3 \left(G+\alpha_\Lambda X^3\right)}{2 n^3}, \,\,\,\,\,\, K_{12} = \frac{3 M^2_{Pl} k^4 \lambda r V F'^2}{N H^2 U W}, \qquad
\\
K_{22} &=& M^2_{Pl}\left(\frac{k^4 a^3 F'}{4 N U} - \frac{k^8 r^2 F'^2}{m^2_g a N H^2 U^2 W}\right), \,\,\,\,\,\, \det{K_{IJ}} = - \frac{9 M^2_{Pl} m^2_g k^4 a^4 \lambda^2 V^2 F'^3}{4 N^2 H^2 U W},
\end{eqnarray}
where
\begin{eqnarray}
U &=& \left(3 F' - \frac{2 k^2}{m^2_g a^2}\frac{(1 + r)}{X}\right), \,\,\,\,\,\, V = \left(1 + \frac{a H \phi H_\phi}{\lambda f}\right), \nonumber
\\
W &=& \Bigg(12 A + 2\left(\frac{2 k^4 r^2}{m^2_g a^4 H^2 U} + 6 X + \frac{3 X^2 r H_\phi}{H}\right)F' + \frac{3 m^2_g X^2 r F'^2}{H^2} - 6 X^2\left(1 - \frac{X r H_\phi}{H}\right)F''\Bigg).\nonumber
\\
\end{eqnarray}

\end{document}